\documentclass[%
reprint,
superscriptaddress,
amsmath,amssymb,
aps,
pra,
]{revtex4-1}
\pdfoutput = 1
\usepackage{physics}
\usepackage{float}
\usepackage{braket}
\usepackage{graphicx}
\usepackage{dcolumn}
\usepackage{bm}

\begin{document}
\title{Out-of-Time-Ordered Correlation in Anisotropic Dicke Model}

\author{Jihan Hu}
\affiliation{Department of Modern Physics, University of Science and Technology of China, Hefei, China}
\author{Shaolong Wan}
\affiliation{Department of Modern Physics, University of Science and Technology of China, Hefei, China}

\date{\today}

\begin{abstract}
Out-of-time-ordered correlation (OTOC) functions have been used as an indicator of quantum chaos in a lot of physical systems. In this work, we computationally demonstrate that zero-temperature OTOC can detect quantum phase transition in anisotropic Dicke model. Phase diagram is given with OTOC. Finite-size effect is studied. Finally, temperature effect is discussed.

\end{abstract}
\maketitle


\section{Introduction}
There has been a great revival of out-of-time-ordered correlation (OTOC) functions in recent years. Its importance is first realized by A.~Larkin and Y.~N.~Ovchinnikov \cite{Larkin1969}. After several decades of silence, its relevance with black holes is pointed out by A.~Y.~Kitaev \cite{Kitaev2014,Kitaev2015}. Consider two generic unitary operators $ V $ and $ W $, along with a many-body Hamiltonian $ H $ of the system. The OTOC is defined as
\begin{equation}
F(t) = \frac{1}{2} \left( \langle V^\dagger(0) W^\dagger(t) V(0) W(t) \rangle_\beta + \text{h.c.} \right),
\label{eq: F}
\end{equation}
where $ V(0)=V $ and $ W(t) = e^{iHt} W e^{-iHt} $ denote time-evolving Hermitian operators. $ \langle...\rangle $ stands for the expectation value on a pure state of interest (in our case the ground state), or the thermal average for a given temperature. The OTOC is closely related to the squared commutator of $ V $ and $ W(t) $, which is written as $ C(t)=\langle|[W(t),V]|^2\rangle = 2(1-\Re F(t)) $. Naively substituting $ W=V=\hat{p} $, we have \cite{Rozenbaum2017}
\begin{equation}\label{eq:C}
\hspace{-5pt} C(t) = \hbar^2\left<\left(\dfrac{\partial\hat{p}(t)}{\partial x(0)}\right)^2 \right> \approx \hbar^2\left<\hspace{-5.83pt}\left<\left(\dfrac{\Delta p(t)}{\Delta x(0)}\right)^2 \right>\hspace{-5.83pt}\right> = C_{\rm cl}(t),
\end{equation}
Where $ \hat{p} $ and $ \hat{x} $ are the momentum and position operators, $ C_{\rm cl}(t) $ is the classical counterpart of $ C(t) $. Then, we can see the intrinsic relation of $ C(t) $ with classical chaos, which attracts a lot of interest \cite{Shenker2014,Shenker2014a,Shenker2015,Maldacena2016}. Over years of studies, OTOC is proved to be powerful in different scenarios including many-body localization \cite{Khemani2017,Chen2017}, information scrambling \cite{Hosur2016} and AdS/CFT \cite{Maldacena2016,Maldacena2016a,Maldacena2016b}.

Theoretical studies on OTOC going on \cite{Blake2018,YungerHalpern2017,Zhang2018,Rammensee2018,Cotler2017}, direct computations show that it can be used as an order parameter to distinguish different quantum phases. There have been flourishing works on Bose-Hubbard model \cite{Shen2017}, Ising model \cite{Heyl2018} and XXZ model \cite{Dag2019}. Even in topological phase transition, OTOC has a substantial footprint \cite{Dag2020}.

The Dicke model is definitely a fundamental model of cavity-QED, describing the interaction of many atoms to a single cavity mode\cite{Emary2003,Emary2003a,Garraway2011,Kirton2018}. This model undergoes a phase transition to a superradiant state at a critical value of the atom-field interaction \cite{Emary2003a}. Classical chaos methods are widely applied to this model even before the recent enthusiasm \cite{Emary2003,Emary2003a,Altland2012,Perez-Fernandez2011,Bastarrachea_Magnani_2015,Chavez-Carlos2016,Chavez-Carlos2019}. In Ref~\cite{Alavirad2019}, diagrammatic method on OTOC is used to compute the Lyapunov exponent and study the chaotic behavior of Dicke model. The interplay of quantum phase transition and chaos remains an open question. Using OTOC as a complementary tool to show ergodic-nonergodic transition in a generalized version of Dicke model, Ref~\cite{Buijsman2017} concludes the existence of a quantum analogue of the classical Kolmogorov-Arnold-Moser (KAM) theorem \cite{Hose1983,Brandino2015}, which might be misleading because the OTOC does not show its power in detecting quantum phase transition.

In this work, we compute OTOC in anisotropic Dicke model. Since the rigid quantum phase transition only occurs at zero temperature, we focus on zero-temperature OTOC. Here, we recover the phase diagram of anisotropic Dicke model with OTOC and study the finite-size effect. Finally, we give the dynamical pattern of OTOC with temperature raised.

\section{The model and the main result}
The anisotropic Dicke model(ADM) can be written as 
	\begin{align} 
	\begin{split}
	H  =& \hbar \omega a^\dagger a + \hbar \omega_0 J_z  + \frac{g_1}{\sqrt{2j}} \left( a^\dagger J_- + a J_+ \right) \\
	& + \frac{g_2}{\sqrt{2j}} \left( a^\dagger J_+ + a J_- \right),
	\end{split} 
	\label{eq: H}
	\end{align}
where $ a $($ a^\dagger $)  are bosonic annihilation (creation) operators, satisfying $ [a,a^\dagger] = 1 $ and  $ J_{\pm,z}=\sum_{i=1}^{2j} \frac{1}{2} \sigma^{(i)}_{\pm,z} $ are angular momentum operators, describing a pseudospin of length $ j $ composed of $ N=2j $ non-interacting spin-$1/2$ atoms described by the Pauli matrices $ \sigma_{\pm, z}^{(i)} $ acting on site $ i $. The ADM describes a single bosonic mode (often a cavity photon mode) of frequency $ \omega $ which interacts collectively with a set of $ N $ two-level systems (the atoms) with energy-splitting $ \omega_z $, within the dipole approximation coupled to the field. Written in terms of collective operators, the ADM can be greatly simplified when we take $ j $ to have its maximal value $ j=N/2 $. The model has four tunable parameters: the photon frequency $ \omega $, the atomic energy splitting $ \omega_z $ and counter-(co-)rotating photon-atom coupling $ g_1 $($ g_2 $).
For $g_1 = g_2 = g$, the ADM reduces to the Dicke model with coupling parameter $ g $. The ADM possess a parity symmetry $ \Pi = \exp(i \pi [a^\dagger a + J_z + j]) $ satisfying $ [H, \Pi] = 0 $ with eigenvalues $ \pm 1 $. 

Our focus is restricted to the positive parity subspace, which includes the ground state for the parameter ranges considered in this work.
Hereafter, we work in the basis $ \{ |n \rangle \otimes |j, m \rangle \} $ with $ a^\dagger a |n \rangle = n |n \rangle $ and $ J_z |j, m \rangle = m |j, m \rangle $ and we set $ \omega=\omega_z=1 $ which is most physically acceptable. We take the cutoff of number of bosons to be 100 unless otherwise stated (i.e. $ n_{\rm{cutoff}}=100 $).

When $ g_1 = 0 $ or $ g_2 = 0 $, the ADM is integrable, which inspires a lot of researchers working on the integrablity of the model. In the thermodynamic limit $ N \to \infty $, the ADM exhibits a second-order quantum phase transition at $ g_1 + g_2 = \sqrt{\omega \omega_z} $ with order parameter $ a^\dagger a / j $ \cite{Emary2003a}, separating the normal phase at $ g_1 + g_2 < \sqrt{\omega \omega_z} $ with $ \langle a^\dagger a\rangle / j = 0 $ from the superradiant phase with $ \langle a^\dagger a \rangle / j = \mathcal{O}(1) $. 

\begin{figure}
	\centering
	\includegraphics[width=\columnwidth]{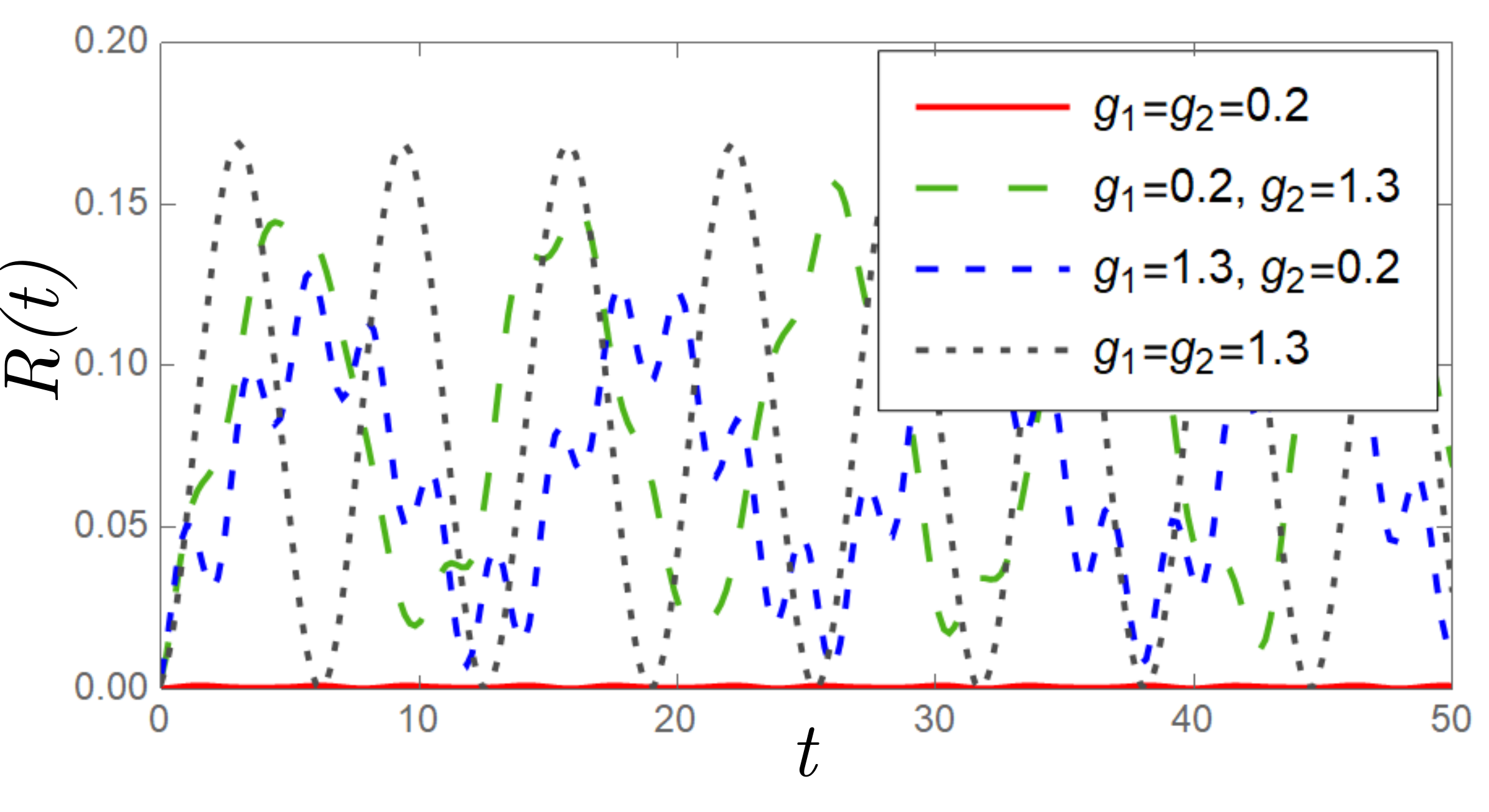}
	\caption{The representative time evolution of $ R(t) $ (defined in the text).}
	\label{fig:tplot}
\end{figure}

\begin{figure}
	\centering
	\includegraphics[width=\columnwidth]{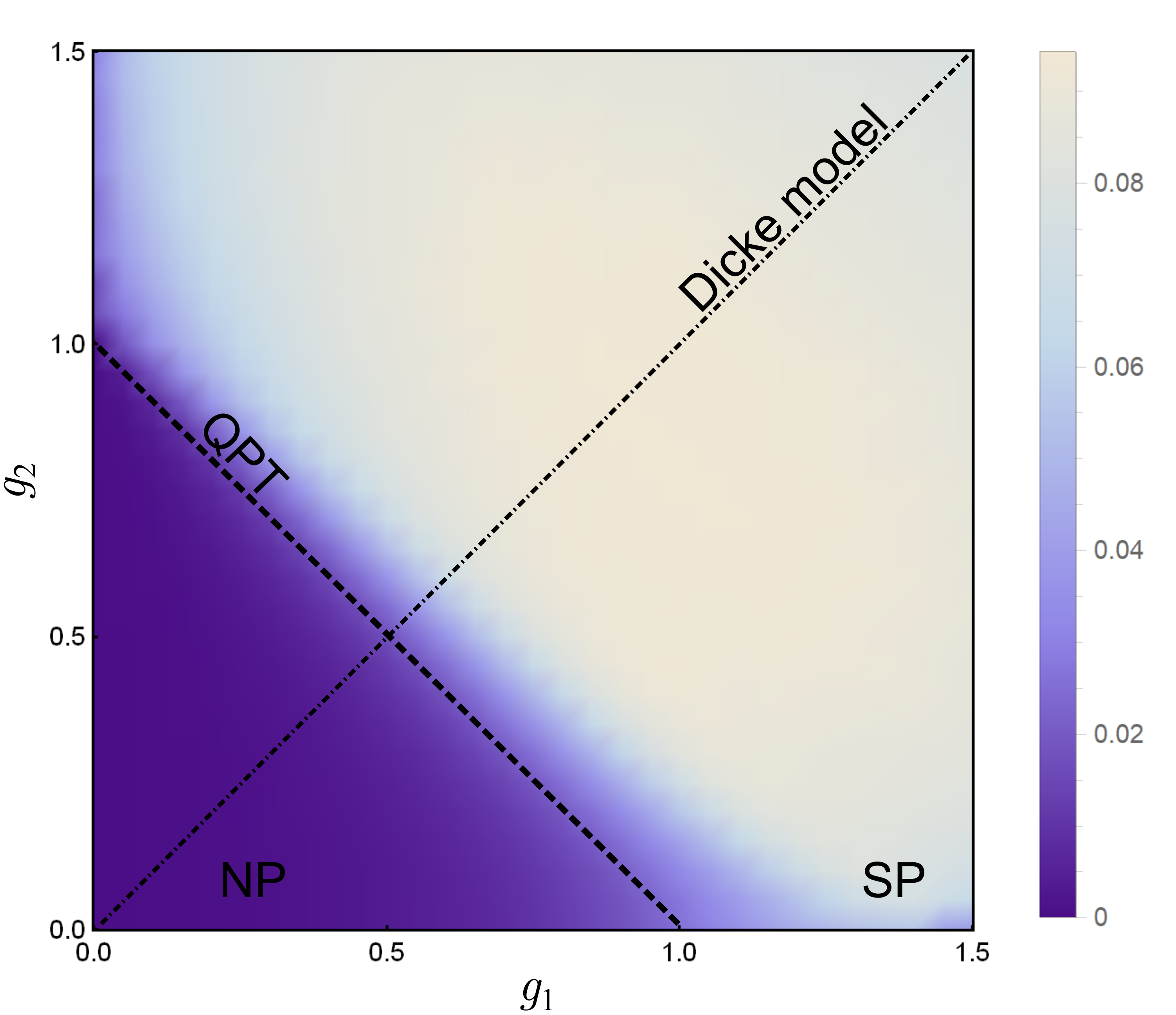}
	\caption{Density plot of $ \bar{\mathcal{R}} $ as a function of $ g_{1,2} $. The Dicke model, i.e. where $ g_1=g_2 $, is indicated by a dot-dashed line. Normal phase (NP) and superradiant phase (SP) are separated by the quantum phase transition (QPT) line, the dashed line.}
	\label{fig:0t}
\end{figure}

In the following, we utilize OTOC to detect the phase transition in the ADM. We take $ W=V= a^\dagger a + 10 $ and compute the OTOC at zero-temperature, i.e. $ F(t)=\langle V^\dagger W^\dagger(t) V W(t)\rangle $, where $ \langle...\rangle $ stands for the expectation value on ground state. We focus on $ R(t)=1-F(t)/F(0) $. This value, dubbed residue OTOC, is large if the system spreads information fast. For a generic system, it is generally close to zero. In Fig.~\ref{fig:tplot}, $R(t)$ are plotted for typical $ g_{1,2} $, as a function of time, from which we can see the OTOC of ADM shows steady behavior. 
This enables us to utilize $ \bar{\mathcal{R}}=\lim_{t\rightarrow \infty}\int_{0}^{t}  R(t') dt' $, which is considered as saturation value\cite{Dag2019}, to separate different phases. As indicated in Fig.~\ref{fig:0t}, in normal phase (blue region),  $ \bar{\mathcal{R}} $ is smaller than in superradiant phase (white region), as a function of $ g_{1,2} $. In general, it is easy to understand that $ F(t) $ will not go far from $ F(0) $, if the interaction between atoms and field is small, leading to a small value of $ \bar{\mathcal{R}} $. We take about $ 30 \times 30 $ points resulting in the saw-tooth pattern of the separatrix, which suggests the existence of a phase boundary.

\section{Finite-size effect and temperature going high}
\begin{figure}
	\centering
	\includegraphics[width=\columnwidth]{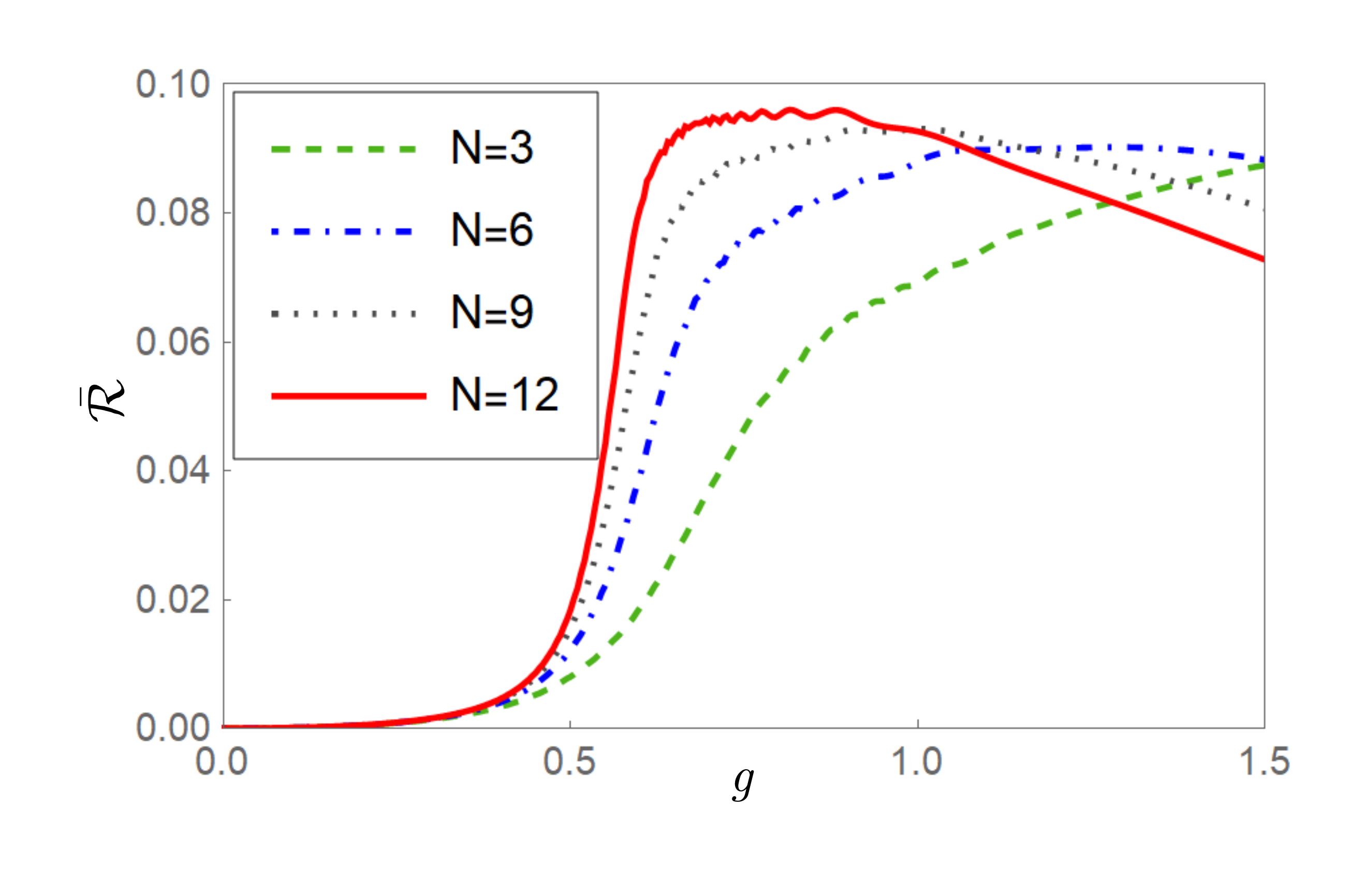}
	\caption{$ \bar{\mathcal{R}} $ as a function of Dicke coupling constant $ g $, plotted with different $ N $ (the size of the system).}
	\label{fig:dicke}
\end{figure}

\begin{figure}
	\centering
	\includegraphics[width=\columnwidth]{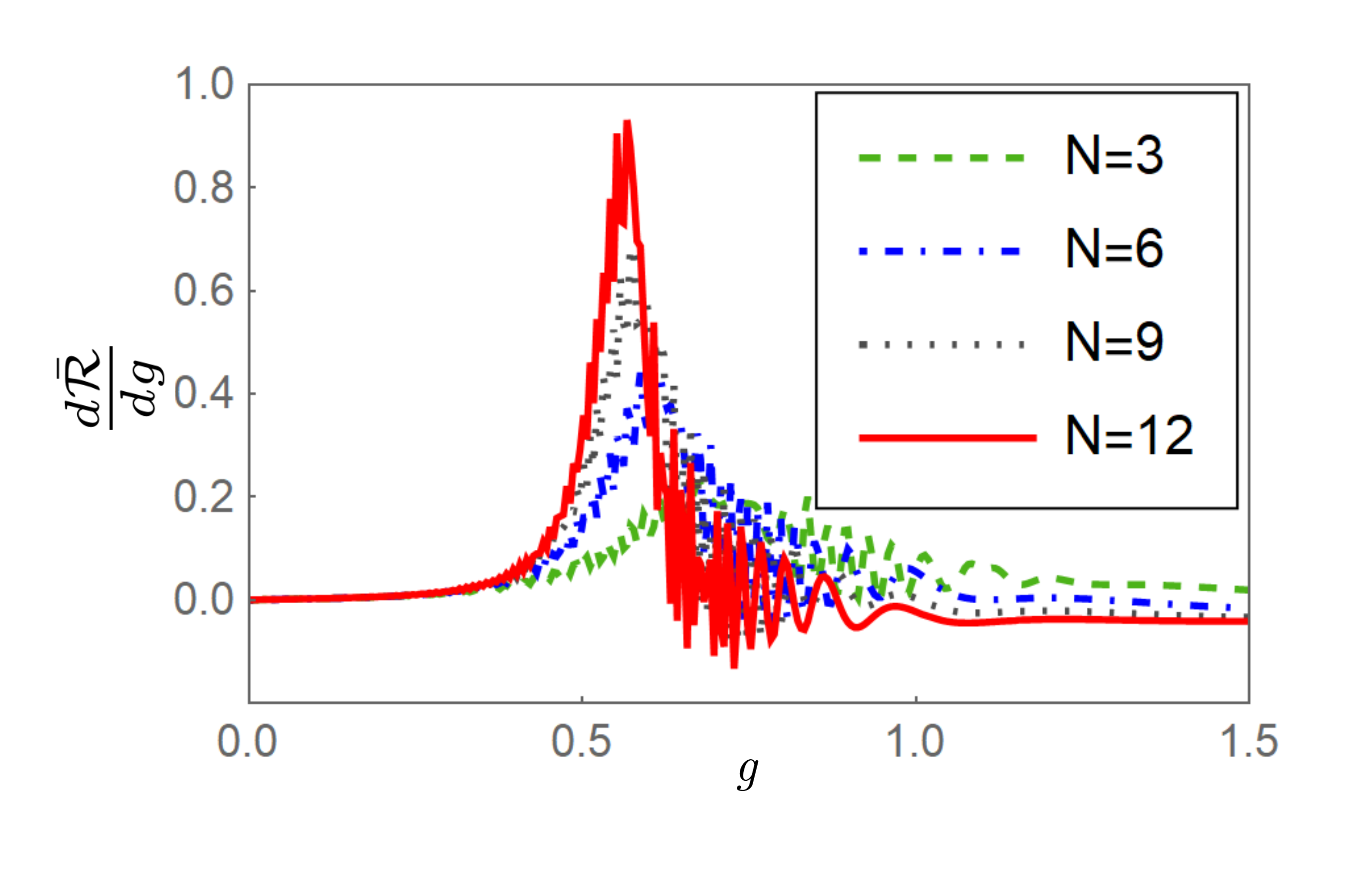}
	\caption{$ \frac{d \bar{\mathcal{R}}}{d g} $ plotted as $ g $. The rapid oscillations are consequences of finite time window we taken.}
	\label{fig:dRdg}
\end{figure}

\begin{figure}
	\centering
	\includegraphics[width=\columnwidth]{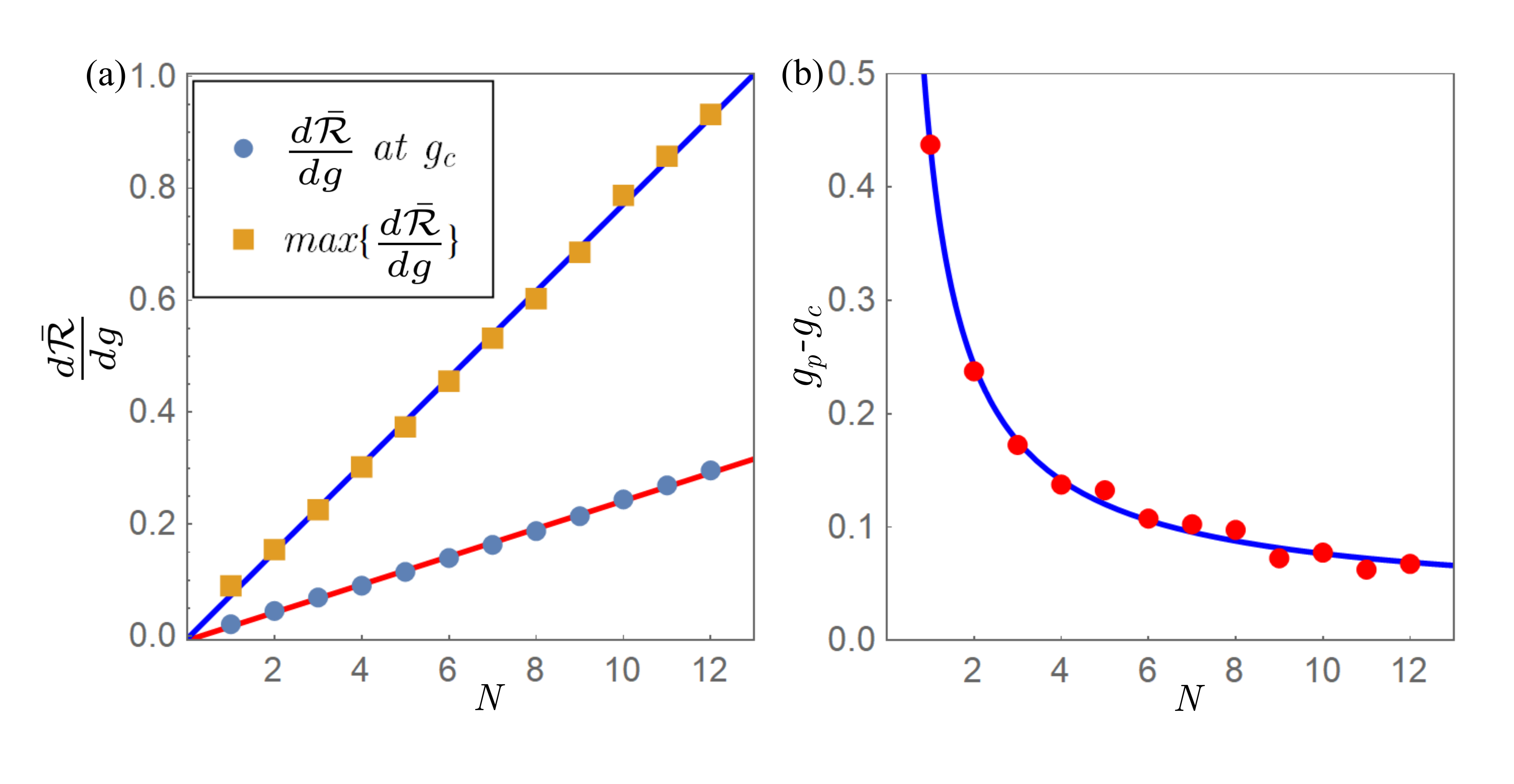}
	\caption{(a) Fittings of $ \frac{d \bar{\mathcal{R}}}{d g} $ at $ g_c=0.5 $ and of maximum of $ \frac{d \bar{\mathcal{R}}}{d g} $, respectively as a function of $ N $. (b) Peak position $ g_p $ minus $ g_c $ as a function of $ N $, tending to $ 0 $ with $ N $ large. Fitting models and parameters are given in the text.}
	\label{fig:fitting}
\end{figure}

To analyze the finite-size effect with $ N $ increasing, we further calculate $ \bar{\mathcal{R}} $ with different $ N $ along the Dicke line as a function of $ g $. Fig.~\ref{fig:dicke} shows close to the critical point($ g \approx 0.5 $), the slope is steeper with N larger. In the thermodynamic limit, we expect at $ g_c=0.5 $, there will be a sudden jump. Note that when $ N\ge 6 $,there is a decreasing tendency with $ g $ going large. We believe it is caused by relatively small $ n_{\rm{cutoff}} $, which does not have any effect on our discussion. 
We plot the derivative of $ \bar{\mathcal{R}} $ with respect to $ g $ in Fig.~\ref{fig:dRdg}. Despite of the rapid oscillating behavior, we can infer there is a QPT near $ g_c=0.5 $. 
To provide more compelling evidence, we give three fittings with $ N $ increasing. In Fig.~\ref{fig:fitting}(a), both the value at $ g_c $ and the maximum of $ \frac{d \bar{\mathcal{R}}}{d g} $ grow linearly with $ N $, fitting models being $ -0.0071+0.0249 N $ (red line) and $ -0.0046+0.0776 N $ (blue line) respectively. In Fig.~\ref{fig:fitting}(b), we fit the data using $ a N^{-b} + c $, with $ a=0.4079 $, $ b=0.9220 $ and $ c=0.0253 $.
The fittings clearly show the existence of a jump of $ \bar{\mathcal{R}} $ at $ g_c $, in the thermodynamic limit. 		

It is believed that OTOC can characterize ergodic-nonergodic transition in the ADM \cite{Buijsman2017}, but with a relatively high temperature ($ T=10 $). 
\begin{figure}[h]
	\centering
	\includegraphics[width=\columnwidth]{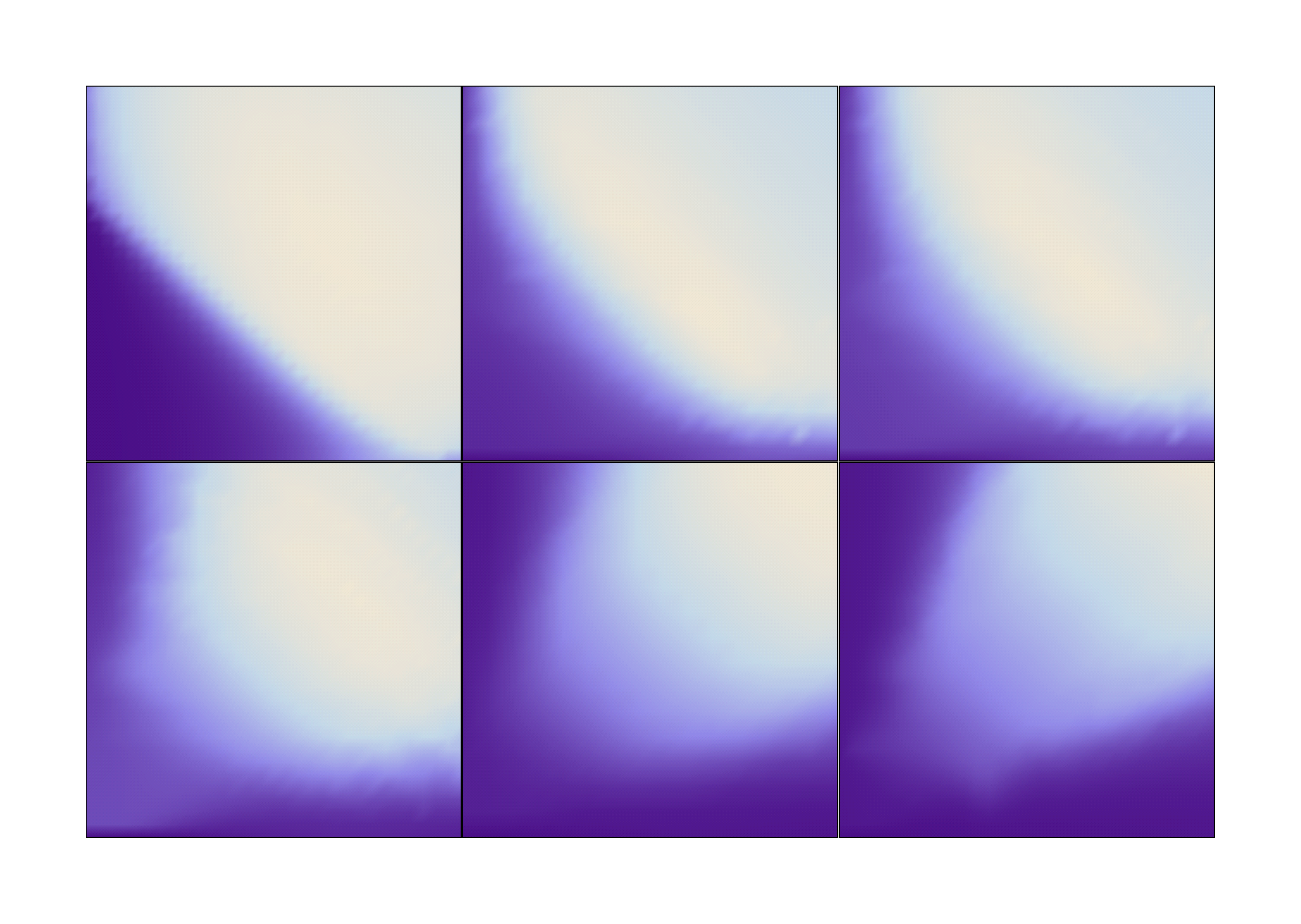}
	\caption{Density plot of $ \bar{\mathcal{R}} $ as a function of $ g_{1,2} $ with $ T=0, 1, 1.43, 3.33, 10, \infty $. Although the legends of each densityplot are different, they do not matter and are left out for simplicity.}
	\label{fig:Tvarying}
\end{figure}
Here, we want to track how a single physical quantity can show quite different physics of system. In Fig.~\ref{fig:Tvarying}, we plot $\bar{\mathcal{R}}$ as a function of $ g_{1,2}$, with $ \beta $ being $ \infty $, $ 1 $, $ 0.7 $, $ 0.3 $, $ 0.1 $, and $ 0 $, viz., temperature being $ 0 $, $ 1 $, $ 1.43 $, $ 3.33 $, $ 10 $ and $ \infty $.
Although, the scalings are different, the patterns are rather smooth, varying from a relatively clear boundary, which shows the existence of QPT, to a relatively blurred one, which is considered as a hint of the quantum KAM theorem. 
	
\section{Conclusion}
In this work, we compute zero-temperature OTOC in the ADM, and recover the phase diagram, which possesses a clear boundary between the normal phase and the superradiant phase. Further finite-size effect is discussed and in the thermodynamic limit, the saturation value of the residue OTOC in Dicke model will be like a step function, which demonstrates that the boundary line in the ADM will be quite clear as the number of atoms going to infinity. We also give a dynamic changing of density plot of $ \bar{\mathcal{R}} $ with temperature raised. At zero-temperature, the boundary is clear and separate superradiant phase from normal phase; at quantitatively high temperature ($ T=10 $), the boundary is fuzzy, which shows an ergodic-nonergodic transition, suggesting the rationality of the quantum KAM theorem. 

\section*{Acknowledgments}

This work was supported by NSFC Grant No.11275180.
	
	\bibliography{dicke_otoc}

\begin{thebibliography}{35}%
\makeatletter
\providecommand \@ifxundefined [1]{%
 \@ifx{#1\undefined}
}%
\providecommand \@ifnum [1]{%
 \ifnum #1\expandafter \@firstoftwo
 \else \expandafter \@secondoftwo
 \fi
}%
\providecommand \@ifx [1]{%
 \ifx #1\expandafter \@firstoftwo
 \else \expandafter \@secondoftwo
 \fi
}%
\providecommand \natexlab [1]{#1}%
\providecommand \enquote  [1]{``#1''}%
\providecommand \bibnamefont  [1]{#1}%
\providecommand \bibfnamefont [1]{#1}%
\providecommand \citenamefont [1]{#1}%
\providecommand \href@noop [0]{\@secondoftwo}%
\providecommand \href [0]{\begingroup \@sanitize@url \@href}%
\providecommand \@href[1]{\@@startlink{#1}\@@href}%
\providecommand \@@href[1]{\endgroup#1\@@endlink}%
\providecommand \@sanitize@url [0]{\catcode `\\12\catcode `\$12\catcode
  `\&12\catcode `\#12\catcode `\^12\catcode `\_12\catcode `\%12\relax}%
\providecommand \@@startlink[1]{}%
\providecommand \@@endlink[0]{}%
\providecommand \url  [0]{\begingroup\@sanitize@url \@url }%
\providecommand \@url [1]{\endgroup\@href {#1}{\urlprefix }}%
\providecommand \urlprefix  [0]{URL }%
\providecommand \Eprint [0]{\href }%
\providecommand \doibase [0]{http://dx.doi.org/}%
\providecommand \selectlanguage [0]{\@gobble}%
\providecommand \bibinfo  [0]{\@secondoftwo}%
\providecommand \bibfield  [0]{\@secondoftwo}%
\providecommand \translation [1]{[#1]}%
\providecommand \BibitemOpen [0]{}%
\providecommand \bibitemStop [0]{}%
\providecommand \bibitemNoStop [0]{.\EOS\space}%
\providecommand \EOS [0]{\spacefactor3000\relax}%
\providecommand \BibitemShut  [1]{\csname bibitem#1\endcsname}%
\let\auto@bib@innerbib\@empty
\bibitem [{\citenamefont {{Larkin}}\ and\ \citenamefont
  {{Ovchinnikov}}(1969)}]{Larkin1969}%
  \BibitemOpen
  \bibfield  {author} {\bibinfo {author} {\bibfnamefont {A.~I.}\ \bibnamefont
  {{Larkin}}}\ and\ \bibinfo {author} {\bibfnamefont {Y.~N.}\ \bibnamefont
  {{Ovchinnikov}}},\ }\href@noop {} {\bibfield  {journal} {\bibinfo  {journal}
  {Soviet Journal of Experimental and Theoretical Physics}\ }\textbf {\bibinfo
  {volume} {28}},\ \bibinfo {pages} {1200} (\bibinfo {year}
  {1969})}\BibitemShut {NoStop}%
\bibitem [{\citenamefont {Kitaev}(2014)}]{Kitaev2014}%
  \BibitemOpen
  \bibfield  {author} {\bibinfo {author} {\bibfnamefont {A.~Y.}\ \bibnamefont
  {Kitaev}},\ }\href@noop {} {\  (\bibinfo {year} {2014})}\BibitemShut
  {NoStop}%
\bibitem [{\citenamefont {Kitaev}(2015)}]{Kitaev2015}%
  \BibitemOpen
  \bibfield  {author} {\bibinfo {author} {\bibfnamefont {A.~Y.}\ \bibnamefont
  {Kitaev}},\ }\href {http:// online.kitp.ucsb.edu/online/entangled15/kitaev/}
  {\  (\bibinfo {year} {2015})}\BibitemShut {NoStop}%
\bibitem [{\citenamefont {Rozenbaum}\ \emph {et~al.}(2017)\citenamefont
  {Rozenbaum}, \citenamefont {Ganeshan},\ and\ \citenamefont
  {Galitski}}]{Rozenbaum2017}%
  \BibitemOpen
  \bibfield  {author} {\bibinfo {author} {\bibfnamefont {E.~B.}\ \bibnamefont
  {Rozenbaum}}, \bibinfo {author} {\bibfnamefont {S.}~\bibnamefont {Ganeshan}},
  \ and\ \bibinfo {author} {\bibfnamefont {V.}~\bibnamefont {Galitski}},\
  }\href {\doibase 10.1103/PhysRevLett.118.086801} {\bibfield  {journal}
  {\bibinfo  {journal} {Phys. Rev. Lett.}\ }\textbf {\bibinfo {volume} {118}},\
  \bibinfo {pages} {086801} (\bibinfo {year} {2017})}\BibitemShut {NoStop}%
\bibitem [{\citenamefont {Shenker}\ and\ \citenamefont
  {Stanford}(2014{\natexlab{a}})}]{Shenker2014}%
  \BibitemOpen
  \bibfield  {author} {\bibinfo {author} {\bibfnamefont {S.~H.}\ \bibnamefont
  {Shenker}}\ and\ \bibinfo {author} {\bibfnamefont {D.}~\bibnamefont
  {Stanford}},\ }\href@noop {} {\bibfield  {journal} {\bibinfo  {journal} {J.
  High Energy Phys.}\ }\textbf {\bibinfo {volume} {3}},\ \bibinfo {pages} {067}
  (\bibinfo {year} {2014}{\natexlab{a}})}\BibitemShut {NoStop}%
\bibitem [{\citenamefont {Shenker}\ and\ \citenamefont
  {Stanford}(2014{\natexlab{b}})}]{Shenker2014a}%
  \BibitemOpen
  \bibfield  {author} {\bibinfo {author} {\bibfnamefont {S.~H.}\ \bibnamefont
  {Shenker}}\ and\ \bibinfo {author} {\bibfnamefont {D.}~\bibnamefont
  {Stanford}},\ }\href@noop {} {\bibfield  {journal} {\bibinfo  {journal} {J.
  High Energy Phys.}\ }\textbf {\bibinfo {volume} {12}},\ \bibinfo {pages}
  {046} (\bibinfo {year} {2014}{\natexlab{b}})}\BibitemShut {NoStop}%
\bibitem [{\citenamefont {Shenker}\ and\ \citenamefont
  {Stanford}(2015)}]{Shenker2015}%
  \BibitemOpen
  \bibfield  {author} {\bibinfo {author} {\bibfnamefont {S.~H.}\ \bibnamefont
  {Shenker}}\ and\ \bibinfo {author} {\bibfnamefont {D.}~\bibnamefont
  {Stanford}},\ }\href@noop {} {\bibfield  {journal} {\bibinfo  {journal} {J.
  High Energy Phys.}\ }\textbf {\bibinfo {volume} {5}},\ \bibinfo {pages} {132}
  (\bibinfo {year} {2015})}\BibitemShut {NoStop}%
\bibitem [{\citenamefont {Maldacena}\ \emph {et~al.}(2016)\citenamefont
  {Maldacena}, \citenamefont {Shenker},\ and\ \citenamefont
  {Stanford}}]{Maldacena2016}%
  \BibitemOpen
  \bibfield  {author} {\bibinfo {author} {\bibfnamefont {J.}~\bibnamefont
  {Maldacena}}, \bibinfo {author} {\bibfnamefont {S.~H.}\ \bibnamefont
  {Shenker}}, \ and\ \bibinfo {author} {\bibfnamefont {D.}~\bibnamefont
  {Stanford}},\ }\href {\doibase 10.1007/JHEP08(2016)106} {\bibfield  {journal}
  {\bibinfo  {journal} {J. High Energy Phys.}\ }\textbf {\bibinfo {volume}
  {8}},\ \bibinfo {pages} {106} (\bibinfo {year} {{2016}})}\BibitemShut
  {NoStop}%
\bibitem [{\citenamefont {Khemani}\ \emph {et~al.}(2018)\citenamefont
  {Khemani}, \citenamefont {Vishwanath},\ and\ \citenamefont
  {Huse}}]{Khemani2017}%
  \BibitemOpen
  \bibfield  {author} {\bibinfo {author} {\bibfnamefont {V.}~\bibnamefont
  {Khemani}}, \bibinfo {author} {\bibfnamefont {A.}~\bibnamefont {Vishwanath}},
  \ and\ \bibinfo {author} {\bibfnamefont {D.~A.}\ \bibnamefont {Huse}},\
  }\href {http://arxiv.org/abs/1710.09835} {\bibfield  {journal} {\bibinfo
  {journal} {Phys. Rev. X}\ }\textbf {\bibinfo {volume} {8}},\ \bibinfo {pages}
  {031057} (\bibinfo {year} {2018})}\BibitemShut {NoStop}%
\bibitem [{\citenamefont {Chen}\ \emph {et~al.}(2017)\citenamefont {Chen},
  \citenamefont {Zhou}, \citenamefont {Huse},\ and\ \citenamefont
  {Fradkin}}]{Chen2017}%
  \BibitemOpen
  \bibfield  {author} {\bibinfo {author} {\bibfnamefont {X.}~\bibnamefont
  {Chen}}, \bibinfo {author} {\bibfnamefont {T.}~\bibnamefont {Zhou}}, \bibinfo
  {author} {\bibfnamefont {D.~A.}\ \bibnamefont {Huse}}, \ and\ \bibinfo
  {author} {\bibfnamefont {E.}~\bibnamefont {Fradkin}},\ }\href
  {https://onlinelibrary.wiley.com/doi/abs/10.1002/andp.201600332} {\bibfield
  {journal} {\bibinfo  {journal} {Ann. Phys.}\ }\textbf {\bibinfo {volume}
  {529}},\ \bibinfo {pages} {1600332} (\bibinfo {year} {2017})}\BibitemShut
  {NoStop}%
\bibitem [{\citenamefont {Hosur}\ \emph {et~al.}(2016)\citenamefont {Hosur},
  \citenamefont {Qi}, \citenamefont {Roberts},\ and\ \citenamefont
  {Yoshida}}]{Hosur2016}%
  \BibitemOpen
  \bibfield  {author} {\bibinfo {author} {\bibfnamefont {P.}~\bibnamefont
  {Hosur}}, \bibinfo {author} {\bibfnamefont {X.-L.}\ \bibnamefont {Qi}},
  \bibinfo {author} {\bibfnamefont {D.~A.}\ \bibnamefont {Roberts}}, \ and\
  \bibinfo {author} {\bibfnamefont {B.}~\bibnamefont {Yoshida}},\ }\href
  {https://doi.org/10.1007/JHEP02(2016)004} {\bibfield  {journal} {\bibinfo
  {journal} {J. High Energy Phys.}\ }\textbf {\bibinfo {volume} {2}},\ \bibinfo
  {pages} {004} (\bibinfo {year} {2016})}\BibitemShut {NoStop}%
\bibitem [{\citenamefont {Maldacena}\ and\ \citenamefont
  {Stanford}(2016)}]{Maldacena2016a}%
  \BibitemOpen
  \bibfield  {author} {\bibinfo {author} {\bibfnamefont {J.}~\bibnamefont
  {Maldacena}}\ and\ \bibinfo {author} {\bibfnamefont {D.}~\bibnamefont
  {Stanford}},\ }\href {\doibase 10.1103/PhysRevD.94.106002} {\bibfield
  {journal} {\bibinfo  {journal} {Phys. Rev. D}\ }\textbf {\bibinfo {volume}
  {94}},\ \bibinfo {pages} {106002} (\bibinfo {year} {2016})}\BibitemShut
  {NoStop}%
\bibitem [{\citenamefont {Maldacena}\ \emph {et~al.}()\citenamefont
  {Maldacena}, \citenamefont {Stanford},\ and\ \citenamefont
  {Yang}}]{Maldacena2016b}%
  \BibitemOpen
  \bibfield  {author} {\bibinfo {author} {\bibfnamefont {J.}~\bibnamefont
  {Maldacena}}, \bibinfo {author} {\bibfnamefont {D.}~\bibnamefont {Stanford}},
  \ and\ \bibinfo {author} {\bibfnamefont {Z.}~\bibnamefont {Yang}},\
  }\href@noop {} {\ }\Eprint {http://arxiv.org/abs/1606.01857v2}
  {arXiv:1606.01857v2} \BibitemShut {NoStop}%
\bibitem [{\citenamefont {Blake}\ \emph {et~al.}(2018)\citenamefont {Blake},
  \citenamefont {Lee},\ and\ \citenamefont {Liu}}]{Blake2018}%
  \BibitemOpen
  \bibfield  {author} {\bibinfo {author} {\bibfnamefont {M.}~\bibnamefont
  {Blake}}, \bibinfo {author} {\bibfnamefont {H.}~\bibnamefont {Lee}}, \ and\
  \bibinfo {author} {\bibfnamefont {H.}~\bibnamefont {Liu}},\ }\href
  {https://doi.org/10.1007/JHEP10(2018)127} {\bibfield  {journal} {\bibinfo
  {journal} {J. High Energy Phys.}\ }\textbf {\bibinfo {volume} {10}},\
  \bibinfo {pages} {127} (\bibinfo {year} {2018})}\BibitemShut {NoStop}%
\bibitem [{\citenamefont {{Yunger Halpern}}(2017)}]{YungerHalpern2017}%
  \BibitemOpen
  \bibfield  {author} {\bibinfo {author} {\bibfnamefont {N.}~\bibnamefont
  {{Yunger Halpern}}},\ }\href@noop {} {\bibfield  {journal} {\bibinfo
  {journal} {Phys. Rev. A}\ }\textbf {\bibinfo {volume} {95}},\ \bibinfo
  {pages} {012120} (\bibinfo {year} {2017})}\BibitemShut {NoStop}%
\bibitem [{\citenamefont {Zhang}\ \emph {et~al.}(2019)\citenamefont {Zhang},
  \citenamefont {Huang},\ and\ \citenamefont {Chen}}]{Zhang2018}%
  \BibitemOpen
  \bibfield  {author} {\bibinfo {author} {\bibfnamefont {Y.-L.}\ \bibnamefont
  {Zhang}}, \bibinfo {author} {\bibfnamefont {Y.}~\bibnamefont {Huang}}, \ and\
  \bibinfo {author} {\bibfnamefont {X.}~\bibnamefont {Chen}},\ }\href
  {https://arxiv.org/pdf/1802.04492.pdf} {\bibfield  {journal} {\bibinfo
  {journal} {Phys. Rev. B}\ }\textbf {\bibinfo {volume} {99}},\ \bibinfo
  {pages} {014303} (\bibinfo {year} {2019})}\BibitemShut {NoStop}%
\bibitem [{\citenamefont {Rammensee}\ \emph {et~al.}(2018)\citenamefont
  {Rammensee}, \citenamefont {Urbina},\ and\ \citenamefont
  {Richter}}]{Rammensee2018}%
  \BibitemOpen
  \bibfield  {author} {\bibinfo {author} {\bibfnamefont {J.}~\bibnamefont
  {Rammensee}}, \bibinfo {author} {\bibfnamefont {J.~D.}\ \bibnamefont
  {Urbina}}, \ and\ \bibinfo {author} {\bibfnamefont {K.}~\bibnamefont
  {Richter}},\ }\href {https://doi.org/10.1103/PhysRevLett.121.124101}
  {\bibfield  {journal} {\bibinfo  {journal} {Phys. Rev. Lett.}\ }\textbf
  {\bibinfo {volume} {121}},\ \bibinfo {pages} {124101} (\bibinfo {year}
  {2018})}\BibitemShut {NoStop}%
\bibitem [{\citenamefont {Cotler}\ \emph {et~al.}(2017)\citenamefont {Cotler},
  \citenamefont {Hunter-Jones}, \citenamefont {Liu},\ and\ \citenamefont
  {Yoshida}}]{Cotler2017}%
  \BibitemOpen
  \bibfield  {author} {\bibinfo {author} {\bibfnamefont {J.}~\bibnamefont
  {Cotler}}, \bibinfo {author} {\bibfnamefont {N.}~\bibnamefont
  {Hunter-Jones}}, \bibinfo {author} {\bibfnamefont {J.}~\bibnamefont {Liu}}, \
  and\ \bibinfo {author} {\bibfnamefont {B.}~\bibnamefont {Yoshida}},\ }\href
  {https://doi.org/10.1007/JHEP11(2017)048} {\bibfield  {journal} {\bibinfo
  {journal} {J. High Energy Phys.}\ }\textbf {\bibinfo {volume} {11}},\
  \bibinfo {pages} {048} (\bibinfo {year} {2017})}\BibitemShut {NoStop}%
\bibitem [{\citenamefont {Shen}\ \emph {et~al.}(2017)\citenamefont {Shen},
  \citenamefont {Zhang}, \citenamefont {Fan},\ and\ \citenamefont
  {Zhai}}]{Shen2017}%
  \BibitemOpen
  \bibfield  {author} {\bibinfo {author} {\bibfnamefont {H.}~\bibnamefont
  {Shen}}, \bibinfo {author} {\bibfnamefont {P.}~\bibnamefont {Zhang}},
  \bibinfo {author} {\bibfnamefont {R.}~\bibnamefont {Fan}}, \ and\ \bibinfo
  {author} {\bibfnamefont {H.}~\bibnamefont {Zhai}},\ }\href@noop {} {\bibfield
   {journal} {\bibinfo  {journal} {Phys. Rev. B}\ }\textbf {\bibinfo {volume}
  {96}},\ \bibinfo {pages} {054503} (\bibinfo {year} {2017})},\ \Eprint
  {http://arxiv.org/abs/1608.02438} {arXiv:1608.02438} \BibitemShut {NoStop}%
\bibitem [{\citenamefont {Heyl}(2018)}]{Heyl2018}%
  \BibitemOpen
  \bibfield  {author} {\bibinfo {author} {\bibfnamefont {M.}~\bibnamefont
  {Heyl}},\ }\href@noop {} {\bibfield  {journal} {\bibinfo  {journal} {Phys.
  Rev. Lett}\ }\textbf {\bibinfo {volume} {121}},\ \bibinfo {pages} {016801}
  (\bibinfo {year} {2018})}\BibitemShut {NoStop}%
\bibitem [{\citenamefont {Daǧ}\ \emph {et~al.}(2019)\citenamefont {Daǧ},
  \citenamefont {Sun},\ and\ \citenamefont {Duan}}]{Dag2019}%
  \BibitemOpen
  \bibfield  {author} {\bibinfo {author} {\bibfnamefont {C.~B.}\ \bibnamefont
  {Daǧ}}, \bibinfo {author} {\bibfnamefont {K.}~\bibnamefont {Sun}}, \ and\
  \bibinfo {author} {\bibfnamefont {L.~M.}\ \bibnamefont {Duan}},\ }\href@noop
  {} {\bibfield  {journal} {\bibinfo  {journal} {Phys. Rev. Lett.}\ }\textbf
  {\bibinfo {volume} {123}},\ \bibinfo {pages} {140602} (\bibinfo {year}
  {2019})},\ \Eprint {http://arxiv.org/abs/1902.05041} {arXiv:1902.05041}
  \BibitemShut {NoStop}%
\bibitem [{\citenamefont {Daǧ}\ \emph {et~al.}(2020)\citenamefont {Daǧ},
  \citenamefont {Duan},\ and\ \citenamefont {Sun}}]{Dag2020}%
  \BibitemOpen
  \bibfield  {author} {\bibinfo {author} {\bibfnamefont {C.~B.}\ \bibnamefont
  {Daǧ}}, \bibinfo {author} {\bibfnamefont {L.~M.}\ \bibnamefont {Duan}}, \
  and\ \bibinfo {author} {\bibfnamefont {K.}~\bibnamefont {Sun}},\ }\href
  {\doibase 10.1103/PhysRevB.101.104415} {\bibfield  {journal} {\bibinfo
  {journal} {Phys. Rev. B}\ }\textbf {\bibinfo {volume} {101}},\ \bibinfo
  {pages} {104415} (\bibinfo {year} {2020})},\ \Eprint
  {http://arxiv.org/abs/1906.05241} {arXiv:1906.05241} \BibitemShut {NoStop}%
\bibitem [{\citenamefont {Emary}\ and\ \citenamefont
  {Brandes}(2003{\natexlab{a}})}]{Emary2003}%
  \BibitemOpen
  \bibfield  {author} {\bibinfo {author} {\bibfnamefont {C.}~\bibnamefont
  {Emary}}\ and\ \bibinfo {author} {\bibfnamefont {T.}~\bibnamefont
  {Brandes}},\ }\href@noop {} {\bibfield  {journal} {\bibinfo  {journal} {Phys.
  Rev. Lett.}\ }\textbf {\bibinfo {volume} {90}},\ \bibinfo {pages} {044101}
  (\bibinfo {year} {2003}{\natexlab{a}})}\BibitemShut {NoStop}%
\bibitem [{\citenamefont {Emary}\ and\ \citenamefont
  {Brandes}(2003{\natexlab{b}})}]{Emary2003a}%
  \BibitemOpen
  \bibfield  {author} {\bibinfo {author} {\bibfnamefont {C.}~\bibnamefont
  {Emary}}\ and\ \bibinfo {author} {\bibfnamefont {T.}~\bibnamefont
  {Brandes}},\ }\href {\doibase 10.1103/PhysRevE.67.066203} {\bibfield
  {journal} {\bibinfo  {journal} {Phys. Rev. E}\ }\textbf {\bibinfo {volume}
  {67}},\ \bibinfo {pages} {066203} (\bibinfo {year}
  {2003}{\natexlab{b}})}\BibitemShut {NoStop}%
\bibitem [{\citenamefont {Garraway}(2011)}]{Garraway2011}%
  \BibitemOpen
  \bibfield  {author} {\bibinfo {author} {\bibfnamefont {B.~M.}\ \bibnamefont
  {Garraway}},\ }\href@noop {} {\bibfield  {journal} {\bibinfo  {journal}
  {Phil. Trans. R. Soc. A}\ }\textbf {\bibinfo {volume} {369}},\ \bibinfo
  {pages} {1137–1155} (\bibinfo {year} {2011})}\BibitemShut {NoStop}%
\bibitem [{\citenamefont {Kirton}\ \emph {et~al.}(2019)\citenamefont {Kirton},
  \citenamefont {Roses}, \citenamefont {Keeling},\ and\ \citenamefont
  {Dalla~Torre}}]{Kirton2018}%
  \BibitemOpen
  \bibfield  {author} {\bibinfo {author} {\bibfnamefont {P.}~\bibnamefont
  {Kirton}}, \bibinfo {author} {\bibfnamefont {M.~M.}\ \bibnamefont {Roses}},
  \bibinfo {author} {\bibfnamefont {J.}~\bibnamefont {Keeling}}, \ and\
  \bibinfo {author} {\bibfnamefont {E.~G.}\ \bibnamefont {Dalla~Torre}},\
  }\href {\doibase 10.1002/qute.201800043} {\bibfield  {journal} {\bibinfo
  {journal} {Advanced Quantum Technologies}\ }\textbf {\bibinfo {volume} {2}},\
  \bibinfo {pages} {1800043} (\bibinfo {year} {2019})}\BibitemShut {NoStop}%
\bibitem [{\citenamefont {Altland}\ and\ \citenamefont
  {Haake}(2012)}]{Altland2012}%
  \BibitemOpen
  \bibfield  {author} {\bibinfo {author} {\bibfnamefont {A.}~\bibnamefont
  {Altland}}\ and\ \bibinfo {author} {\bibfnamefont {F.}~\bibnamefont
  {Haake}},\ }\href@noop {} {\bibfield  {journal} {\bibinfo  {journal} {Phys.
  Rev. Lett.}\ }\textbf {\bibinfo {volume} {108}},\ \bibinfo {pages} {073601}
  (\bibinfo {year} {2012})}\BibitemShut {NoStop}%
\bibitem [{\citenamefont {P{\'{e}}rez-Fern{\'{a}}ndez}\ \emph
  {et~al.}(2011)\citenamefont {P{\'{e}}rez-Fern{\'{a}}ndez}, \citenamefont
  {Rela{\~{n}}o}, \citenamefont {Arias}, \citenamefont {Cejnar}, \citenamefont
  {Dukelsky},\ and\ \citenamefont {Garc{\'{i}}a-Ramos}}]{Perez-Fernandez2011}%
  \BibitemOpen
  \bibfield  {author} {\bibinfo {author} {\bibfnamefont {P.}~\bibnamefont
  {P{\'{e}}rez-Fern{\'{a}}ndez}}, \bibinfo {author} {\bibfnamefont
  {A.}~\bibnamefont {Rela{\~{n}}o}}, \bibinfo {author} {\bibfnamefont {J.~M.}\
  \bibnamefont {Arias}}, \bibinfo {author} {\bibfnamefont {P.}~\bibnamefont
  {Cejnar}}, \bibinfo {author} {\bibfnamefont {J.}~\bibnamefont {Dukelsky}}, \
  and\ \bibinfo {author} {\bibfnamefont {J.~E.}\ \bibnamefont
  {Garc{\'{i}}a-Ramos}},\ }\href@noop {} {\bibfield  {journal} {\bibinfo
  {journal} {Phys. Rev. E}\ }\textbf {\bibinfo {volume} {83}},\ \bibinfo
  {pages} {046208} (\bibinfo {year} {2011})}\BibitemShut {NoStop}%
\bibitem [{\citenamefont {Bastarrachea-Magnani}\ \emph
  {et~al.}(2015)\citenamefont {Bastarrachea-Magnani}, \citenamefont {del
  Carpio}, \citenamefont {Lerma-Hern{\'{a}}ndez},\ and\ \citenamefont
  {Hirsch}}]{Bastarrachea_Magnani_2015}%
  \BibitemOpen
  \bibfield  {author} {\bibinfo {author} {\bibfnamefont {M.~A.}\ \bibnamefont
  {Bastarrachea-Magnani}}, \bibinfo {author} {\bibfnamefont {B.~L.}\
  \bibnamefont {del Carpio}}, \bibinfo {author} {\bibfnamefont
  {S.}~\bibnamefont {Lerma-Hern{\'{a}}ndez}}, \ and\ \bibinfo {author}
  {\bibfnamefont {J.~G.}\ \bibnamefont {Hirsch}},\ }\href {\doibase
  10.1088/0031-8949/90/6/068015} {\bibfield  {journal} {\bibinfo  {journal}
  {Physica Scripta}\ }\textbf {\bibinfo {volume} {90}},\ \bibinfo {pages}
  {068015} (\bibinfo {year} {2015})}\BibitemShut {NoStop}%
\bibitem [{\citenamefont {Ch\'avez-Carlos}\ \emph {et~al.}(2016)\citenamefont
  {Ch\'avez-Carlos}, \citenamefont {Bastarrachea-Magnani}, \citenamefont
  {Lerma-Hern\'andez},\ and\ \citenamefont {Hirsch}}]{Chavez-Carlos2016}%
  \BibitemOpen
  \bibfield  {author} {\bibinfo {author} {\bibfnamefont {J.}~\bibnamefont
  {Ch\'avez-Carlos}}, \bibinfo {author} {\bibfnamefont {M.~A.}\ \bibnamefont
  {Bastarrachea-Magnani}}, \bibinfo {author} {\bibfnamefont {S.}~\bibnamefont
  {Lerma-Hern\'andez}}, \ and\ \bibinfo {author} {\bibfnamefont {J.~G.}\
  \bibnamefont {Hirsch}},\ }\href {\doibase 10.1103/PhysRevE.94.022209}
  {\bibfield  {journal} {\bibinfo  {journal} {Phys. Rev. E}\ }\textbf {\bibinfo
  {volume} {94}},\ \bibinfo {pages} {022209} (\bibinfo {year}
  {2016})}\BibitemShut {NoStop}%
\bibitem [{\citenamefont {Ch\'avez-Carlos}\ \emph {et~al.}(2019)\citenamefont
  {Ch\'avez-Carlos}, \citenamefont {L\'opez-del Carpio}, \citenamefont
  {Bastarrachea-Magnani}, \citenamefont {Str\'ansk\'y}, \citenamefont
  {Lerma-Hern\'andez}, \citenamefont {Santos},\ and\ \citenamefont
  {Hirsch}}]{Chavez-Carlos2019}%
  \BibitemOpen
  \bibfield  {author} {\bibinfo {author} {\bibfnamefont {J.}~\bibnamefont
  {Ch\'avez-Carlos}}, \bibinfo {author} {\bibfnamefont {B.}~\bibnamefont
  {L\'opez-del Carpio}}, \bibinfo {author} {\bibfnamefont {M.~A.}\ \bibnamefont
  {Bastarrachea-Magnani}}, \bibinfo {author} {\bibfnamefont {P.}~\bibnamefont
  {Str\'ansk\'y}}, \bibinfo {author} {\bibfnamefont {S.}~\bibnamefont
  {Lerma-Hern\'andez}}, \bibinfo {author} {\bibfnamefont {L.~F.}\ \bibnamefont
  {Santos}}, \ and\ \bibinfo {author} {\bibfnamefont {J.~G.}\ \bibnamefont
  {Hirsch}},\ }\href {\doibase 10.1103/PhysRevLett.122.024101} {\bibfield
  {journal} {\bibinfo  {journal} {Phys. Rev. Lett.}\ }\textbf {\bibinfo
  {volume} {122}},\ \bibinfo {pages} {024101} (\bibinfo {year}
  {2019})}\BibitemShut {NoStop}%
\bibitem [{\citenamefont {Alavirad}\ and\ \citenamefont
  {Lavasani}(2019)}]{Alavirad2019}%
  \BibitemOpen
  \bibfield  {author} {\bibinfo {author} {\bibfnamefont {Y.}~\bibnamefont
  {Alavirad}}\ and\ \bibinfo {author} {\bibfnamefont {A.}~\bibnamefont
  {Lavasani}},\ }\href {\doibase 10.1103/PhysRevA.99.043602} {\bibfield
  {journal} {\bibinfo  {journal} {Phys. Rev. A}\ }\textbf {\bibinfo {volume}
  {99}},\ \bibinfo {pages} {043602} (\bibinfo {year} {2019})}\BibitemShut
  {NoStop}%
\bibitem [{\citenamefont {Buijsman}\ \emph {et~al.}(2017)\citenamefont
  {Buijsman}, \citenamefont {Gritsev},\ and\ \citenamefont
  {Sprik}}]{Buijsman2017}%
  \BibitemOpen
  \bibfield  {author} {\bibinfo {author} {\bibfnamefont {W.}~\bibnamefont
  {Buijsman}}, \bibinfo {author} {\bibfnamefont {V.}~\bibnamefont {Gritsev}}, \
  and\ \bibinfo {author} {\bibfnamefont {R.}~\bibnamefont {Sprik}},\ }\href
  {\doibase 10.1103/PhysRevLett.118.080601} {\bibfield  {journal} {\bibinfo
  {journal} {Phys. Rev. Lett.}\ }\textbf {\bibinfo {volume} {118}},\ \bibinfo
  {pages} {080601} (\bibinfo {year} {2017})}\BibitemShut {NoStop}%
\bibitem [{\citenamefont {Hose}\ and\ \citenamefont {Taylor}(1983)}]{Hose1983}%
  \BibitemOpen
  \bibfield  {author} {\bibinfo {author} {\bibfnamefont {G.}~\bibnamefont
  {Hose}}\ and\ \bibinfo {author} {\bibfnamefont {H.~S.}\ \bibnamefont
  {Taylor}},\ }\href {\doibase 10.1103/PhysRevLett.51.947} {\bibfield
  {journal} {\bibinfo  {journal} {Phys. Rev. Lett.}\ }\textbf {\bibinfo
  {volume} {51}},\ \bibinfo {pages} {947} (\bibinfo {year} {1983})}\BibitemShut
  {NoStop}%
\bibitem [{\citenamefont {Brandino}\ \emph {et~al.}(2015)\citenamefont
  {Brandino}, \citenamefont {Caux},\ and\ \citenamefont
  {Konik}}]{Brandino2015}%
  \BibitemOpen
  \bibfield  {author} {\bibinfo {author} {\bibfnamefont {G.~P.}\ \bibnamefont
  {Brandino}}, \bibinfo {author} {\bibfnamefont {J.-S.}\ \bibnamefont {Caux}},
  \ and\ \bibinfo {author} {\bibfnamefont {R.~M.}\ \bibnamefont {Konik}},\
  }\href {\doibase 10.1103/PhysRevX.5.041043} {\bibfield  {journal} {\bibinfo
  {journal} {Phys. Rev. X}\ }\textbf {\bibinfo {volume} {5}},\ \bibinfo {pages}
  {041043} (\bibinfo {year} {2015})}\BibitemShut {NoStop}%
\end{thebibliography}%
	
\end{document}